\documentclass[twocolumn,amsmath,amssymb,prl,superscriptaddress,aps]{revtex4-1}

\usepackage{graphicx}
\usepackage{grffile} 

\usepackage{bm}
\usepackage{bbm}
\usepackage{subfigure}
\usepackage{color}
\usepackage[utf8]{inputenc}
\usepackage{amsmath}

\newcommand{\ket}[1]{\left| \hspace{0.2ex} #1 \right\rangle} 

\begin{document}
\title{Rydberg-Rydberg interaction profile from the excitation dynamics of ultracold atoms in lattices}

\date{\today}

\author{Michael Mayle}
\affiliation{JILA, University of Colorado and National Institute of Standards and Technology, Boulder, Colorado 80309-0440, USA}

\author{Wolfgang Zeller}
\affiliation{Zentrum für Optische Quantentechnologien, Universität 
Hamburg, Luruper Chaussee 149, 22761 Hamburg, Germany}

\author{Nikolas Tezak}
\affiliation{Zentrum für Optische Quantentechnologien, Universität 
Hamburg, Luruper Chaussee 149, 22761 Hamburg, Germany}

\author{Peter Schmelcher}
\affiliation{Zentrum für Optische Quantentechnologien, Universität 
Hamburg, Luruper Chaussee 149, 22761 Hamburg, Germany}

\date{\today}

\begin{abstract}
We propose a method for the determination of the interaction potential of Rydberg atoms. Specifically, we consider a laser-driven Rydberg gas confined in a one-dimensional lattice and demonstrate that the Rydberg atom number after a laser excitation cycle as a function of the laser detuning provides a measure for the Rydberg interaction coefficient. With the lattice spacing precisely known, the proposed scheme only relies on the measurement of the number of Rydberg atoms and thus circumvents the necessity to map the interaction potential by varying the interparticle separation.
\end{abstract}

\maketitle

Among many fascinating systems encountered in modern ultracold atomic and molecular physics are Rydberg atoms, i.e., highly excited atoms with large principal quantum number $n$, whose size can easily exceed that of ground state atoms by several orders of magnitude \cite{gallagher94}. The enormous displacement of the atomic charges makes Rydberg atoms highly susceptible to external fields and at the same time is the origin of their strong mutual interaction. In ultracold gases, the strong dipole-dipole interaction has been shown theoretically \cite{dipoleblockadetheoretically} and experimentally \cite{dipoleblockadeexperimentally} to entail a blockade mechanism in the laser excitation of Rydberg atoms, thereby effectuating a collective excitation process of Rydberg atoms \cite{collectiveexcitation}. Two recent experiments even demonstrated the blockade between two single atoms a few micrometers apart \cite{twoatomblockade}. The strong dipole-dipole interaction renders Rydberg atoms also promising candidates for the implementation of protocols realizing quantum gates or efficient multiparticle entanglement \cite{RevModPhys.82.2313}. In fact, only very recently a {\sc cnot} gate between two individually addressed neutral atoms and the generation of entanglement has been demonstrated experimentally by employing the Rydberg blockade mechanism \cite{twoatomentanglement}. Collective excitation of Rydberg atoms is currently being studied extensively \cite{Sun2008,olmos:043419}, predicting amongst others the dynamical creation of crystalline photonic states \cite{PhysRevLett.104.043002}, the formation of fermionic collective excitations in a lattice gas of Rydberg atoms \cite{PhysRevLett.103.185302}, or the ability to act as a universal quantum simulator \cite{Weimer2010}.

The key ingredient for many of the predicted effects in ultracold Rydberg physics is the strong and long-ranged mutual interaction of Rydberg atoms which has predominantly van der Waals character at large separations \cite{RevModPhys.82.2313}. Hence, the precise knowledge of the corresponding interaction potentials is crucial for making reliable predictions as well as for correctly interpreting experimental data. Accordingly, the long-range Rydberg interaction has been discussed in several recent works and theoretical predictions were made \cite{0953-4075-38-2-021,interactions}. Experimentally, the mechanical effect of van der Waals interactions has been observed in an ultracold Rydberg gas \cite{amthor:023004}.

In this letter we propose a scheme to determine the interaction strength between two Rydberg atoms by means of the Rydberg blockade mechanism. Specifically, we consider the coherent Rydberg excitation of a one-dimensional lattice of trapped ground state atoms. A thorough analysis of the excitation spectrum allows us to identify distinctive values for the detuning of the excitation laser from the single atom resonance that entail a well defined number of Rydberg atoms within the lattice. From these detunings the interaction strength of two Rydberg atoms can be deduced. The possibility to determine the interaction strength of two Rydberg atoms by investigating laser-driven samples of ground state atoms was first mentioned in Ref.~\cite{ates:023002} where the anti-blockade of Rydberg excitation is investigated. Experimentally, this anti-blockade effect was observed in \cite{PhysRevLett.104.013001}. However, the experiment was conducted in a frozen gas rather than in a lattice and was thus not aimed to predict the interaction strength of Rydberg atoms.

\begin{figure}
\includegraphics[width=8cm]{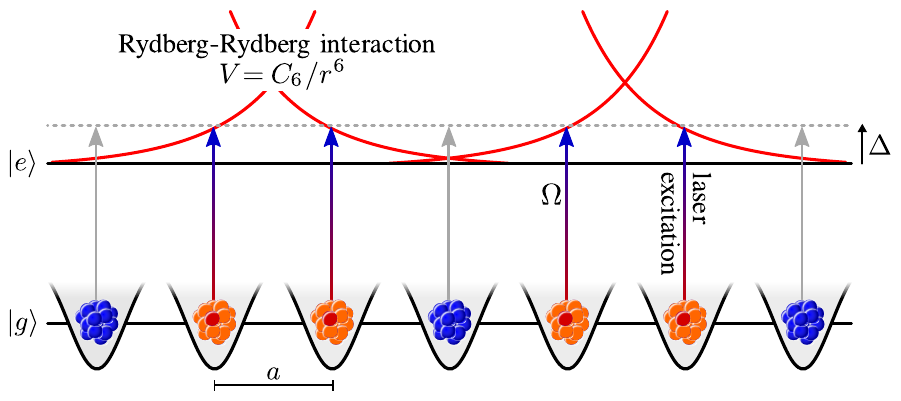}
 \caption{Schematics of the considered setup. Ultracold atoms in a lattice are excited from the ground state $\ket{g}$ (blue/dark gray) to their Rydberg state $\ket{e}$ (red/light gray). The Rydberg-Rydberg interaction shifts the excitation from its single atom resonance (top of the figure). This dipole blockade effect can be compensated by the laser detuning $\Delta$. In this particular example the excitation is resonant to the many-body state $\ket{geegeeg}$.
\label{fig:overview}}
\end{figure}

As in Ref.~\cite{Tezak2010}, we consider an ultracold gas of ground state atoms in a one-dimensional lattice consisting of $N$ sites with regular spacing $a$ and a site-independent filling of $N_0$ atoms. The lattice is subject to a laser excitation to the Rydberg state with a single atom Rabi frequency $\Omega_0$. While this laser transition is usually of two-photon character, the involved intermediate state can be removed from the state space by an adiabatic elimination if the excitation lasers are strongly detuned with respect to this level \cite{0953-4075-43-15-155003}. Moreover, we assume that the dipole blockade is effective within each lattice site $k$ such that our system can be described by means of an ensemble of individual two-level systems, each of which possessing a ground state $\ket{g}_k$ and an excited state $\ket{e}_k$ that are laser-coupled with a collective Rabi frequency $\Omega=\sqrt{N_0}\Omega_0$. Often, these particular two-level systems where the Rydberg excitation is shared among $N_0$ atoms are referred to as \emph{superatoms}. A sketch of our setup is provided in Fig.~\ref{fig:overview}. Applying the rotating wave approximation, one obtains the Hamiltonian \cite{Tezak2010}
\begin{align}\label{eq:master_hamiltonian_constant_couplings}
	H & = \frac{\Omega}{2} \sum_{k=1}^N \sigma_x^{(k)} + \frac{\Delta}{2} \sum_{k=1}^N \sigma_z^{(k)} + V \sum_{l>k}^N \frac{n_e^{(k)}n_e^{(l)}}{|l-k|^m}.
\end{align}
The operators $\sigma^{(k)}_i, i \in \{x,y,z\}$ act only on the superatom located at site $k$ and take on the usual Pauli-matrix form when expressed in the local superatom basis $\left\{\ket{e}_{k},\ket{g}_{k}\right\}$. The excitation number operator reads then $n_e^{(k)} = \frac{1}{2} [\sigma_z^{(k)} + \mathbbm 1]$. Hamiltonian (\ref{eq:master_hamiltonian_constant_couplings}) is valid within the frozen gas limit where no movement of the Rydberg atoms is assumed. For $n=70$ and an initial separation of $a=10\, \mu$m, the change of the distance of two Rydberg atoms due to their repulsive interaction is estimated to be less than a fraction of $10^{-4}$ for an evolution time of $1\,\mu$s \cite{PhysRevLett.104.043002}; this displacement increases for smaller initial distances and higher principal quantum numbers (we consider $^{87}$Rb atoms as an ubiquitous example whenever actual numbers are provided). To actively prevent vast movement of the Rydberg atoms, one might leave the optical trap on during Rydberg excitation. For $nS_{1/2}$ Rydberg states with $n\gtrsim40$, ionization lifetimes of Rydberg atoms due to far infrared trapping lasers are estimated to reach the $\mu$s regime \cite{1367-2630-8-8-163}, which is sufficient for the present purpose. 

Hamiltonian (\ref{eq:master_hamiltonian_constant_couplings}) consists of three different contributions. The laser coupling of each single atom's ground state to the excited state is given by $\frac{1}{2}\Omega \sum_{k=1}^N  \sigma_x^{(k)}$. $\frac{1}{2}\Delta \sum_{k=1}^N  \sigma_z^{(k)}$ provides the energy gap $\Delta$ between the excited state and the ground state of the superatom at each individual site. The interaction strength between two Rydberg atoms at neighboring sites is given by $V=C_m/a^m$ and is considered to be a perturbation of the atomic energy levels. Here, we focus on the regime of a dominant van der Waals interaction ($m=6$); for lattice spacings in the $\mu$m regime, the next order contribution ($m=8$) is more than two orders of magnitude smaller. Moreover, we assume a repulsive interaction, $V>0$, which is common for Rydberg atoms in their $nS_{1/2}$ state \cite{0953-4075-38-2-021}. 

\begin{figure}
\includegraphics[width=8cm]{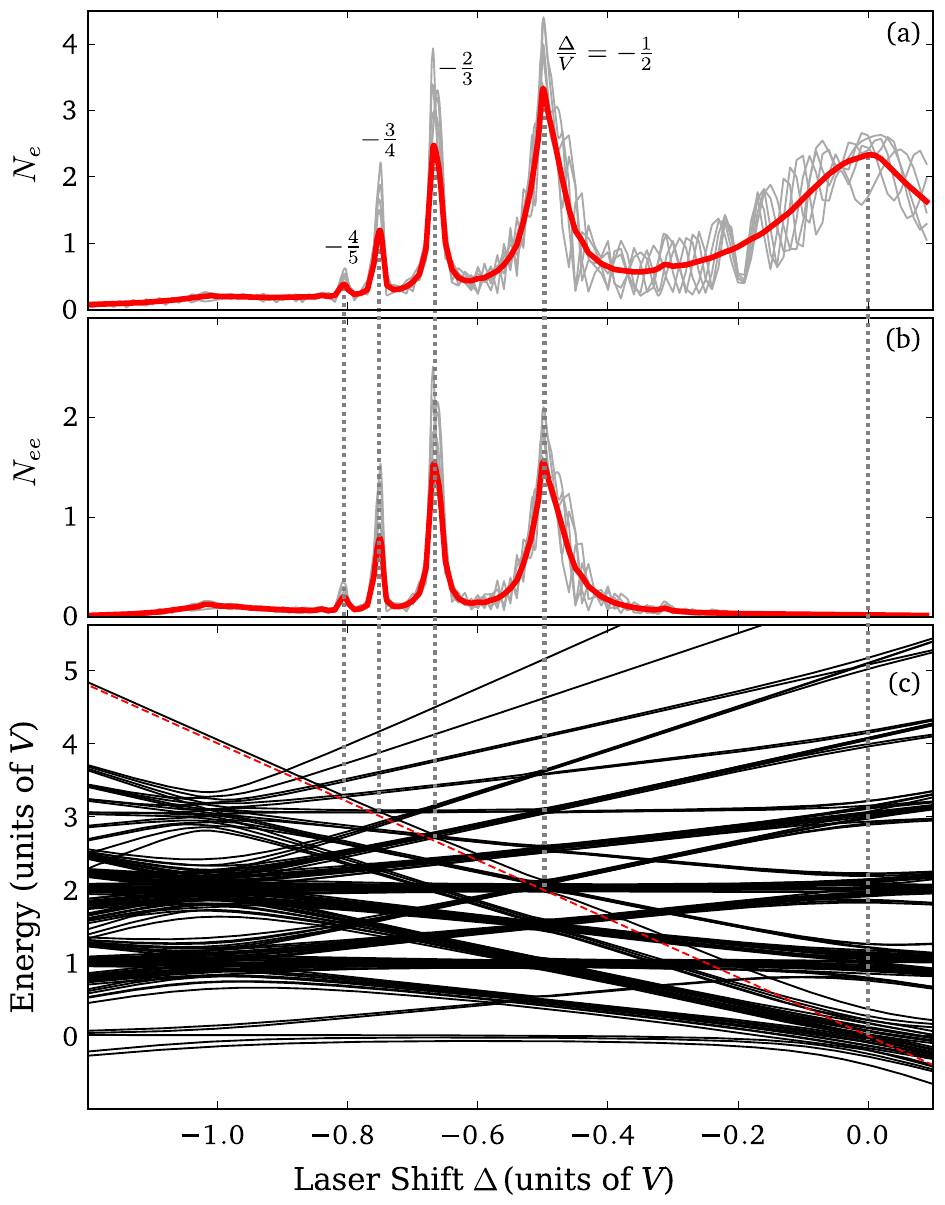}
 \caption{\label{fig:spectrum} (a) Total number of Rydberg atoms and (b) number of neighboring Rydberg excitations after the laser excitation of a $N=8$ lattice of ground state atoms as a function of the detuning from the single atom resonance; a laser coupling strength of $\Omega=0.15\,V$ is considered. The thin gray lines correspond to excitation durations of $T=15,\, 18,\, 21,\, 24,$ and $27\,\Omega^{-1}$. The thick red line is an average over many different laser cycle times. (c) Energy spectrum for the same setup. The dashed line represents the canonical ground state $\ket{G}$ for $\Omega=0$. The peaks of panels (a) and (b) occur at the points where $\ket{G}$ crosses many-body states possessing different number of Rydberg excitations, as described by Eq.~(\ref{eq:deltakappa}).}
\end{figure}

For a weak laser coupling, the contributions diagonal in the spin basis dominate the Hamiltonian. In this case the laser coupling leads to a small off-diagonal perturbation. We thus assume $|\Omega| \ll V, |\Omega| \ll |\Delta| $ and divide the Hamiltonian into two parts, $H = H_0 + H'$, grouping together the laser detuning with the next-neighbor Rydberg interactions to give the dominant contribution
\begin{align}\label{eq:H0}
	H_0  &  = \frac{\Delta}{2} \sum_{k=1}^N \sigma_z^{(k)} + V \sum_{k=1}^{N-1} n_e^{(k)}n_e^{(k+1)},
\end{align}
while the perturbation consists of the laser coupling as well as the long-range Rydberg interactions
\begin{align}\label{eq:Hprime}
	H'  &  = \frac{\Omega}{2} \sum_{k=1}^N \sigma_x^{(k)} + \frac{V}{2^m} \sum_{l = 2}^{N-1} \frac{1}{(l/2)^m } \sum_{k = 1}^{N-l} n_e^{(k)} n_e^{(k+l)}.
\end{align}
Introducing operators for the total excitation number, $ N_e = \sum_{k=1}^N n_e^{(k)}$, and the next-neighbor excitation pair number, $N_{ee} = \sum_{k=1}^{N-1} n_e^{(k)}n_e^{(k+1)}$, we can rewrite the unperturbed Hamiltonian as
\begin{align}
	H_0 & = \Delta \left( N_e - N/2 \right)+ V N_{ee}.
\end{align}
For a given canonical product state $\ket{\alpha} = \ket{s_1s_2\dots s_N}$, $s_k\in\{g_k,e_k\}$ the unperturbed energy eigenvalue is given by
\begin{align}\label{eq:energy_ne_nee}
	E(\alpha) = \Delta [N_e(\alpha) - N/2] + V N_{ee}(\alpha),
\end{align}
where $N_e(\alpha)$ and $N_{ee}(\alpha)$ denote the integral eigenvalues of the operators $N_e$ and $N_{ee}$ for the state $\ket{\alpha}$, respectively. In general, there are multiple canonical product states of equal $(N_e, N_{ee})$ which are degenerate with respect to the unperturbed Hamiltonian. Furthermore, for specific values of the ratio of $\Delta$ and $V$, the simple form of Eq.~\eqref{eq:energy_ne_nee} allows for additional degeneracies between states belonging to different $(N_e, N_{ee})$. Figure~\ref{fig:spectrum}(c) presents the energy eigenvalues of the full Hamiltonian as a function of a varying ratio $\Delta/V$. The points of high degeneracy only occur when $\Delta/V$ takes on certain rational values. For $\Delta=0$ the laser is tuned to atomic resonance. In this case, for $H_0$, the canonical ground state $\ket{G}\equiv\ket{gg\dots g}$ is degenerate with all states that lack neighboring excitations, i.e., dipole-blockaded states. For non-zero detunings, the canonical ground state can also be brought to degeneracy with other states. Physically, this corresponds to a situation in which the lasers resonantly couple $\ket{G}$ to states containing neighboring excitations, compensating the Rydberg-Rydberg interaction by means of the detuning. An example is given by $\Delta/V = -1/2$: In this case $\ket{G}$ is degenerate with states containing exactly two neighboring excitations $\left\{ \ket{eeg\dots g},\ket{geeg\dots g},\dots \ket{g\dots gee}\right\}$. The highest degeneracy, i.e., the minimal number of energy bands, is achieved for $\Delta / V = - 1$.

Here, we are interested in the specific ratios $\Delta/V$ where the state $\ket{G}$ becomes degenerate with states possessing different numbers of Rydberg excitations. Particularly, this occurs at 
\begin{align}\label{eq:deltakappa}
\frac{\Delta_\kappa}{V}=-1+\kappa^{-1},\quad \kappa\ge2,
\end{align}
cf.\ Eq.~(\ref{eq:energy_ne_nee}). At these detunings, the canonical ground state is resonant with the states that satisfy $N_{ee}/N_e=(\kappa-1)/\kappa$, e.g., all states containing $N_e=\kappa \in \mathbb{N}_{\geq2}$ Rydberg excitations whereof $N_{ee}=\kappa-1$ reside at neighboring lattice sites. Performing the Rydberg excitation at such defined detunings will thus drive the lattice in states that contain an increasing number of Rydberg excitation and -pairs. Starting from the canonical ground state at $t=0$, we solved the time-dependent Schr\"odinger equation associated with Hamiltonian (\ref{eq:master_hamiltonian_constant_couplings}). Figure \ref{fig:spectrum}(a) shows the Rydberg excitation number $N_e$ after the excitation cycle ($t=T$) as a function of the laser detuning from the single atom resonance; different durations $T$ of the laser excitation are considered. As expected from Eq.~(\ref{eq:deltakappa}), we find at the well-defined values $\Delta_\kappa$ sharp peaks in the Rydberg atom number $N_e$. The same holds for the number $N_{ee}$ of Rydberg pairs as illustrated in Fig.~\ref{fig:spectrum}(b). From Fig.~\ref{fig:spectrum}(c) it becomes evident that the canonical ground state crosses submanifolds of many-body states at values for the detuning other than specified by Eq.~(\ref{eq:deltakappa}). However, these states are only weakly coupled to $\ket{G}$, giving rise to very narrow avoided crossings. As a result, unrealistically long propagation times are necessary to actually populate these many-body states. Correspondingly, no resonance peaks are visible in Figs.~\ref{fig:spectrum}(a) and (b) at the corresponding detunings.

Relating the detunings $\Delta_\kappa$ to our theoretical predictions allows the determination of the van der Waals interaction coefficient via the relation 
\begin{equation}\label{eq:c6}
C_6=-\frac{\kappa}{\kappa-1}\Delta_\kappa a^6. 
\end{equation}
Since the lattice spacing $a$ is usually known very precisely, any uncertainty in the determination of $C_6$ via Eq.~(\ref{eq:c6}) stems solely from the measurement of the resonant detunings $\Delta_\kappa$. Because of the rather broad $\Delta=0$ peak that serves as reference for the detunings, an accurate determination of the $\Delta/V<0$ resonance positions can be difficult. As a remedy, employing the relative position $\tilde\Delta_{\kappa,\kappa+1}=\Delta_\kappa-\Delta_{\kappa+1}$ of neighboring peaks provides a more reliable source for precision measurements. One finds $C_6=\kappa(\kappa+1)\tilde\Delta_{\kappa,\kappa+1} a^6$. If unknown, the actual value of $\kappa$ can be determined by relating two neighboring peaks, $\Delta_\kappa/\Delta_{\kappa+1}=1-\kappa^{-2}$. Experimentally, $N_e$ can be determined by field ionization of the Rydberg atoms and their subsequent detection via micro-channel plates. For the measurement of $N_{ee}$, a spatially resolved detection is necessary.

The magnitude of the $\Delta_\kappa$ -- and therefore the experimental feasibility -- is determined by the lattice constant $a$ as well as the principal Rydberg quantum number $n$. The expected relevant detunings $\Delta_\kappa$ can be estimated by employing the theoretically determined $C_6$ coefficient of Ref.~\cite{0953-4075-38-2-021}. Taking $a=10\,\mu$m as a typical example for micro-sized dipole or magnetic trap arrays, we find for $n=70$ a separation of the $\kappa=2$ and the $\kappa=3$ resonance peaks in the kilohertz regime, $\tilde\Delta_{2,3}=2\pi\times146\,$kHz. Due to the strong $a^6$ scaling, even larger separations can be achieved for smaller lattice constants: at $a=5\,\mu$m, $\tilde\Delta_{2,3}=2\pi\times9.4\,$MHz. These detunings need to be compared with the finite linewidth of the Rydberg state. For the $ns_{1/2}$ Rydberg states of $^{87}$Rb one finds $\Gamma_n=2\pi\times 0.699\,(n^*)^{-2.94}$\;GHz where $n^*=n-\delta$ is the effective principal quantum number of the Rydberg state including the quantum defect $\delta$ \cite{PhysRevA.65.031401}. For our specific example this yields $\Gamma_{70}=2\pi\times 3.0\,$kHz. Since the latter scales as $n^{-3}$ and $\tilde\Delta_{\kappa,\kappa+1}$ scales as $n^{11}$, the finite  linewidth of the Rydberg state will eventually prevent the resolution of the individual resonance peaks for smaller $n$. Nevertheless, there exists a broad parameter regime that does not put any restrictions on the experimental feasibility.

The actual population of the desired many-body states that contain multiple Rydberg excitations relies on multiphoton processes since -- starting from the canonical ground state -- Hamiltonian (\ref{eq:master_hamiltonian_constant_couplings}) can only couple many-body states that differ exactly by one Rydberg excitation. As a result, the excitation lasers need to be effective long enough for a sufficient population of the desired final state. For the example of Fig.~\ref{fig:spectrum}, we employed a maximal duration of $T_\text{max}=30\,\Omega^{-1}$. For $n=70$ and a lattice constant of $a=5\,\mu$m this yields a time scale of $3.6\,\mu$s which is more than one order of magnitude shorter than the corresponding Rydberg lifetime.
Another possible limitation of the proposed scheme stems from the uncertainty in the filling number $N_0$ of the lattice. Any (site-dependent) variation in this number will affect the excitation dynamics by altering the collective Rabi frequency $\Omega$. However, for long enough excitation times $T$ we do not expect a strong influence on our results since the excitation numbers $N_e$ and $N_{ee}$ are predominantly determined by the detuning $\Delta$ of the laser rather than its coupling strength $\Omega$. We also note that Eq.~(\ref{eq:deltakappa}) is derived from Hamiltonian (\ref{eq:H0}) that neglects any influence of $H'$, cf.\ Eq.~(\ref{eq:Hprime}), on the spectrum. Including the latter slightly shifts [less than 2\% in the case of Fig.~\ref{fig:spectrum}] and splits apart the level crossings, cf.\ Fig.~\ref{fig:spectrum}(c). As a result, the peaks of Figs.~\ref{fig:spectrum}(a) and (b) develop a substructure which, however, can only be resolved for unrealistically long excitation times and a high resolution in $\Delta$. If needed, the numerical determination of this substructure poses no difficulties. Because of this substructure, the widths of the resonances in Fig.~\ref{fig:spectrum} increase for increasing lattice sizes.

While the present work focuses on the excitation of $nS_{1/2}$ Rydberg states, a possible extension of the proposed scheme is the application to higher angular momentum states for which the excitation cannot be modeled by means of a simple two-level system due to numerous possible molecular states with different symmetries. Instead of a single $C_6$ coefficient one could possibly measure an effective interaction strength that averages over all possible molecular symmetries and that is relevant for practical purposes \cite{amthor:023004}. On the other hand, different molecular interaction potential could also lead to separate peaks in the excitation spectrum which renders the extraction of a $C_6$ coefficient ambiguous.

In contrast, for the case of interacting $nS_{1/2}$ Rydberg states one encounters a clean situation: The singlet and the triplet molecular states can be assumed degenerate for the interparticle distances we are considering \cite{0953-4075-38-2-021} and any effects resulting from the hyperfine structure can be diminished by choosing the polarizations of the excitation lasers appropriately. For our example of
$^{87}$Rb, the excitation scheme $5S_{1/2}(F=m_F=2) \leftrightarrow 5P_{3/2}(F=m_F=3) \leftrightarrow nS_{1/2}(F=m_F=2)$ via circularly polarized lasers ensures that only these three hyperfine levels contribute to the excitation process.

To conclude, we have shown that a coherently excited Rydberg gas in a one-dimensional lattice can be used for the determination of the interaction strength of Rydberg atoms. With the lattice spacing precisely known, the proposed scheme only relies on the measurement of the number of Rydberg atoms after an excitation cycle, circumventing the mapping of the interaction energy as a function of distance.

\begin{acknowledgments}
M.M. acknowledges financial support from the German Academic Exchange Service (DAAD).
\end{acknowledgments}

\end{document}